# Nodal gaps in the nematic superconductor FeSe from heat capacity

Frédéric Hardy,[1, *] Mingquan He,[1] Liran Wang,[1] Thomas Wolf,[1] Peter Schweiss,[1] Michael Merz,[1] Maik Barth,[1] Peter Adelmann,[1] Robert Eder,[1] Amir-Abbas Haghighirad,[1,2] and Christoph Meingast[1]

[1]*Institute for Solid-State Physics, Karlsruhe Institute of Technology, 76021 Karlsruhe, Germany*
[2]*Clarendon Laboratory, Department of Physics, University of Oxford,*
*Parks Road, Oxford OX1 3PU, United Kingdom*
(Dated: 20/07/18)

Superconductivity in FeSe has recently attracted a great deal of attention because it emerges out of an electronic nematic state of elusive character. Here we study both the electronic normal state and the superconducting gap structure using heat-capacity measurements on high-quality single crystals. The specific-heat curve, from 0.4 K to $T_c = 9.1$ K, is found to be consistent with a recent gap determination using Bogoliubov quasiparticle interference [P. O. Sprau *et al.*, Science **357**, 75 (2017)], however only if nodes are introduced on either the electron or the hole Fermi-surface sheets. Our analysis, which is consistent with quantum-oscillation measurements, relies on the presence of only two bands, and thus the fate of the theoretically predicted second electron pocket remains mysterious.

In cuprate and iron pnictide superconductors there are strong indications that pairing is magnetically mediated, *i.e.* by the exchange of spin fluctuations [1–3]. A key ingredient to identify the pairing mechanism is to study the momentum dependence of the superconducting gap, $\Delta(\mathbf{k})$. In both materials, $\Delta(\mathbf{k})$ changes sign between $\mathbf{k}$ and $\mathbf{k}' = \mathbf{k} + \mathbf{Q}$, where $\mathbf{Q}$ denotes the momentum at which the spin fluctuation–mediated pairing interaction is peaked. Whereas in cuprates the sign change occurs for $\mathbf{Q} = (\pi, \pi)$ leading to a $d_{x^2-y^2}$ state, $\mathbf{Q}$ connects hole and electron Fermi-surface sheets in pnictides, giving rise to an $s\pm$ state [1–3].

Unlike in pnictides, long-range magnetism is absent in FeSe, although strong magnetic fluctuations exist at low temperature [4]. Here, superconductivity emerges from a nematic state that breaks the $C_4$ symmetry below $T_s \simeq 90$ K. Nematicity manifests itself as a dramatic orbital-dependent shrinking of the Fermi surface [5] in the presence of strong orbital-selective local correlations [6]. Orbital selectivity could also be relevant for superconductivity as first pointed out by Agterberg for $Sr_2RuO_4$ [7], *i.e.* Cooper pairs predominantly binds electrons that share the same orbital character, leading to a highly anisotropic, possibly even nodal order parameter.

Experimentally and theoretically, there is an emerging consensus for extremely anisotropic gaps in FeSe. Using low-temperature Bogoliubov quasiparticle interference and scanning tunneling microscopy (BQPI/STM), Sprau *et al.* [8] found that $\Delta(\mathbf{k})$ is highly anisotropic, but nodeless, on both electron and hole pockets with $\Delta_{max}/\Delta_{min} \gtrsim 15$, more or less consistent with heat-capacity data [9]. The Fermi surface averaged gaps on the electron and hole pockets are of nearly equal magnitude but opposite in sign, *i.e.* $<\Delta_h>_k = 1.5$ meV and $<\Delta_e>_k = -1.2$ meV [8]. Angle-resolved photoemis-sion spectroscopy (ARPES) measurements [10–13] confirm the strong gap anisotropy but are inconclusive about the existence of line nodes due to the slightly higher temperature (T > 1.8 K) of the measurements. Both ARPES [11, 13] and BQPI/STM [8] however agree that $\Delta(\mathbf{k})$ is maximal (minimal) on the portions of the Fermi surface where the $d_{yz}$ ($d_{xz}$ and $d_{xy}$) orbital weight is dominant. These results were explained theoretically either by invoking orbital-selective correlations [8, 14], nematic pairing [15] or nematic pairing from orbital selective spin fluctuations [16]. On the other hand, thermal-transport [17, 18], penetration-depth, [19, 20] and heat-capacity studies, [21–24] all reported essentially a strong two-band behavior with $\delta = \frac{<\Delta_h(k)>_k}{<\Delta_e(k)>_k} \approx 3$ - 10.

In this Letter, we report a detailed thermodynamic study of 7 different batches of vapor-grown FeSe single crystals as a function of temperature and magnetic field. We argue that the Sommerfeld coefficient $\gamma_n$ within the nematic phase is consistent with a Fermi surface that consists of only one hole and one electron bands, instead of the three pockets expected theoretically [14–16]. Excellent agreement of our specific-heat can be obtained using the anisotropic gaps derived from the BQPI/STM data [8, 26], however only if one enforces line nodes on one of the pockets. This nodal behavior is supported by the field dependence of the mixed-state heat capacity. Further, it is found to be surprisingly robust against disorder. Finally, similar to $BaFe_2As_2$ [25] we find an anomalous linear temperature-dependent susceptibility from $T > T_s$ to 700 K, a feature that appears to be a hallmark of the undoped Fe-based superconductors.

Figures 1a-b show the temperature dependence of the resistivity and specific heat of Samples 1 and 2, which are characteristic of our best crystals. The quite large residual resistivity ratio, $RRR \equiv \frac{\rho(300K)}{\rho(0K)} \approx 200$, in com-

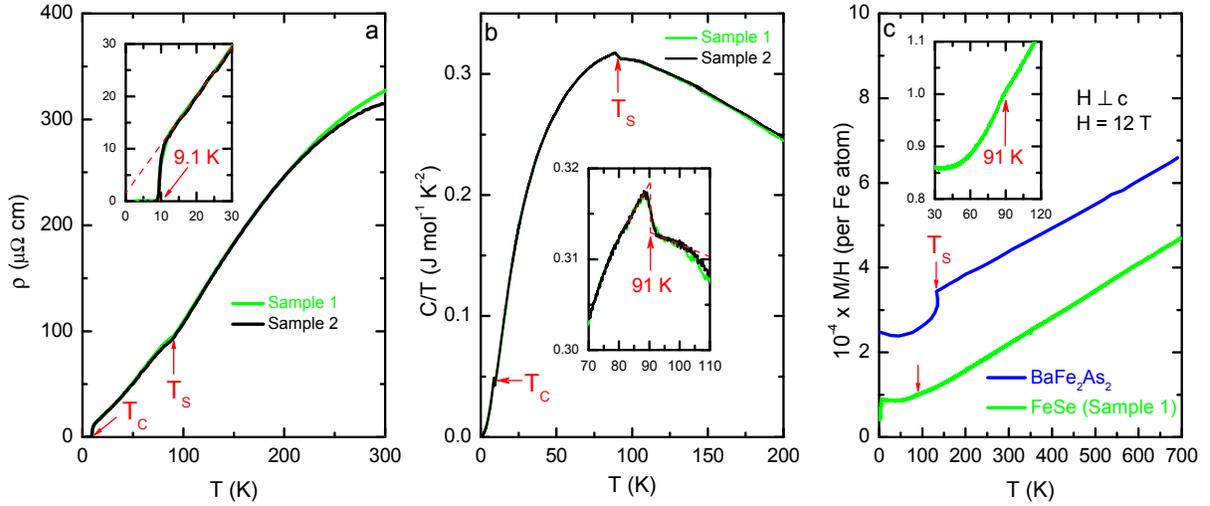

FIG. 1. (a) Resistivity of Samples 1 and 2. The inset shows a magnification of the low temperature region. The dashed line is a linear extrapolation of the normal-state resistivity to T = 0. (b) Heat capacity of Samples 1 and 2. (c) Comparison of the magnetic susceptibility of Sample 1 to that of $BaFe_2As_2$ reproduced from Ref.[25]. In (b) and (c), the insets show data around $T_s$ = 91 K on an enlarged scale.

parison to previous studies [8, 9, 18, 22, 27, 28], is an indication of a high crystal quality. A clear step-like anomaly related to nematicity is observed at $T_s$ = 91 K.

Figure 2a demonstrates that the low-temperature electronic heat capacity $C_e(T)$ of Samples 1 and 2 is practically identical to that of two other single crystals, all grown under similar conditions. We find a sharp transition at $T_c \approx 9.1$ K, with a width of 0.7 K and $\Delta C_e/\gamma_n T_c \approx 1.75$ indicative of weak-to-moderate coupling. Details about the sample preparation and the lattice subtraction are given in Supplementary Notes 1 and 2. Superconductivity is almost fully suppressed for H = 14 T, revealing a temperature-independent $C_e/T$ for $T < 12$ K (see Fig.2a) indicative of a nematic Landau Fermi liquid despite the nearly linear resistivity (see Fig.1a). The Sommerfeld coefficient $\gamma_n \equiv \frac{C_e}{T}$ amounts to 6.9 ± 0.1 mJ mol$^{-1}$ K$^{-2}$ within the nematic state. We note that $\gamma_n$ is quite small and almost equal to that of $BaFe_2As_2$ deep inside the SDW phase [29].

A simple inspection of the data shown in Fig.2a suggests that both electron and hole gaps are comparable in magnitude, since the typical low-temperature hump-like feature of superconductors with both small and large gaps, as found e.g. in $MgB_2$ [30] and $KFe_2As_2$ [31] (see Supplementary Note 3), is missing in the FeSe data. This is consistent with the gaps derived from both BQPI/STM and ARPES [8, 12, 13] but not with the strong two-band character inferred from both penetration-depth, [19] and thermal-conductivity measurements [18]. For $T \leqslant 3$ K, $C_e/T$ decreases linearly with temperature (see inset of Fig.2a) and extrapolates at $T = 0$ to a negligibly small residual density of states $0 \leqslant \gamma_0 < 0.2$ mJ mol$^{-1}$ K$^{-2}$ which amounts to less than 3 % of $\gamma_n$. This linear behavior represents clear evidence for the existence of line nodes in our FeSe crystals. Our data are substantially different from all previous specific-heat measurements. In particular, an excess specific heat was observed for T < 4 K in Refs [9, 22, 24, 28, 32], which was interpreted as a signature of a tiny energy gap. In Supplementary Note 4 we show that a more likely explanation involves a Schottky anomaly from paramagnetic impuritiesand, besides, the 1 K anomaly, reported by Chen et al. [9], is observed only in two of our crystals measured under special conditions and therefore may not represent an intrinsic feature of FeSe.

Further evidence for the existence of nodes can be obtained by examining the field dependence of $\gamma(T, H) \equiv C_e(T, H)/T$ in the mixed state depicted in Fig.2b. Indeed, Volovik [33–36] showed that $\gamma(T, H)$ is proportional to $\gamma_n\sqrt{H}$, for $\frac{T}{T_c} << \sqrt{\frac{H}{H_{c2}}}$, in nodal superconductors due to the Doppler shift experienced by the quasiparticles with momentum near the nodal directions. As illustrated in the inset of Fig.2b, this is the case e.g. for Samples 1 and 5 for $\frac{T}{T_c} < 0.15\sqrt{\frac{H}{H_{c2}}}$, indicative of nodal gaps. On the other hand, for $H/H_{c2} > 0.4$, we find that $\gamma(T, H)$ follows a quasi-linear field dependence, which is anomalous for an orbitally-limited nodal superconductor. As mentioned in Ref.[23], this is most likely related to sizable paramagnetic effects [23, 37, 38], since the Pauli field, $H_p = \frac{\sqrt{2}\Delta}{g\mu_B}$, is estimated to be of the order of 18 T in FeSe.

In the following we make a direct comparison of our heat capacity with that calculated from the BQPI/STM

measurements. We use the data of Sample 2 with a near zero value of $\gamma_0$. In Fig.3a, we reproduce the angular dependence of $\Delta_{h,e}(\phi)$ inferred from the BQPI/STM experiments (solid symbols) [8] together with our fits (solid lines) using the leading angular harmonic approximation (LAHA) [1, 15], as done in Ref.[9]. The resulting heat capacity, computed within the two-band $\alpha$-model [30], is compared to our measurements in Figs 3b-c (red lines). Here, we impose equal density of states on both bands as inferred from the quantum-oscillation data (see Supplementary Note 5), so that in principle there are no fit parameters. $C_e/T$ derived from the BQPI/STM gap structure reproduces our data very well, but only down to $T/T_c \approx 0.3$. Below this temperature, the calculation underestimates the specific heat, reflecting the existence of low-lying nodal excitations in our crystals (see Fig.3c). To improve the agreement at low temperature, we modify our fit of $\Delta_{h,e}(\phi)$ to allow for the possibility of line nodes on either the hole or the electron band, as displayed in Figs 3d and 3g (red areas), respectively. In both cases, the calculation, shown respectively in Figs 3e-f and 3h-i, is now in excellent agreement with our measurements over the entire temperature range. This provides another strong indication for the existence of line nodes in FeSe. This is also consistent with the recent calculations of Benfatto et al. [16], which predict nodes only on the electron pocket. Finally, the possibility of line nodes on both bands appears to be ruled out (see Figs 3j-k-l) since the calculation now overestimates the measured specific heat for $T/T_c < 0.3$.

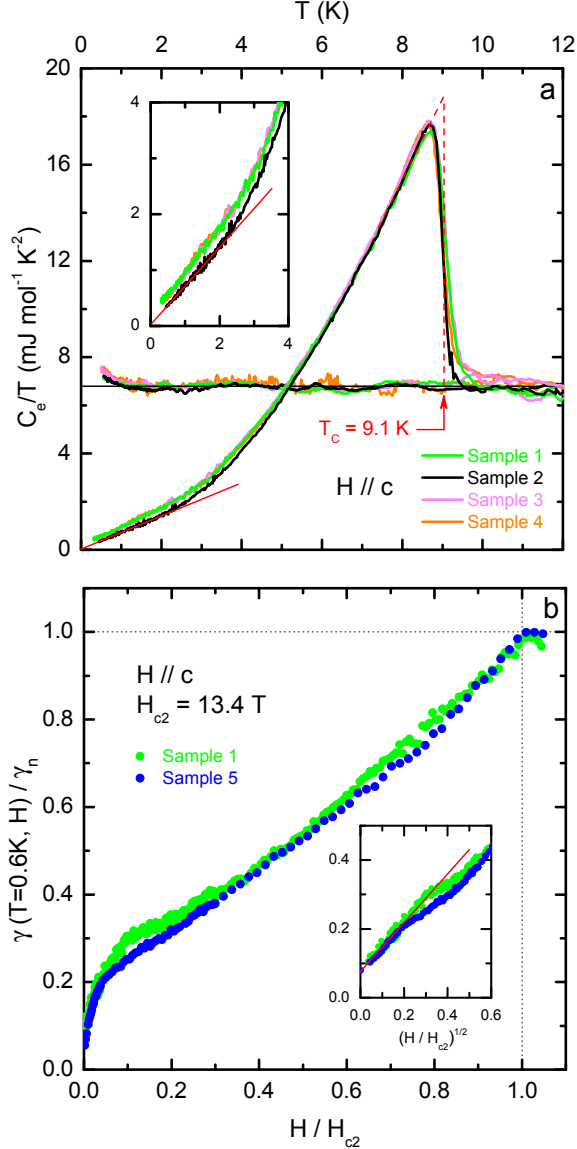

FIG. 2. (a) Low-temperature $C_e(T)$ of Samples 1, 2, 3 and 4 in 0 and 14 T. The dashed line represents an entropy–conserving construction that yields $T_c \approx 9.1$ K. The inset shows data below 4 K on an enlarged scale revealing a linear $C_e/T$ indicative of line nodes (solid line). (b) Mixed-state heat capacity $\gamma(T, H)$ of Samples 1 and 5 at T = 0.6 K. The inset shows $\gamma(T, H)$ as a function of $H^{1/2}$ in the low-field region.

Next, we discuss the role of disorder on the gap structure by comparing $C_e(T)$ of several single crystals grown under different conditions (see Supplementary Note 1). Although no discernible changes in composition could be resolved using x-ray diffraction, significant differences are observed in our specific-heat measurements (see Figure 4). First, a substantial reduction of $T_c$ of about 7 % and 14 % is found for Samples 5 and 6, respectively, indicating that disorder is sizably pair-breaking. This is in line with the observation that scattering by Fe vacancies [39] and twin boundaries [26] produces in-gap bound states in FeSe. Similarly, the normalized specific-heat anomaly, $\frac{\Delta C}{\gamma_n T_c}$, is reduced significantly from 1.73 in Sample 1 to 1.26 in Sample 6, indicating that $<\Delta_{h,e}>_k$ are also reduced by defect scattering. The linear $C_e/T$ at low temperature is quite robust and persists in all samples in spite of the significant reduction of both $T_c$ and $\frac{\Delta C}{\gamma_n T_c}$ (see inset of Fig.4). This result is quite surprising, because significant interband scattering is expected to produce sizable gapless excitations for a sign-changing order parameter [40, 41].

The exact electronic structure in the nematic state of FeSe is still highly disputed. Theoretically one expects to have one hole and two electron bands [8, 14–16] but ARPES spectra however have been interpreted as providing evidence for either one or two electron bands [5, 6, 12, 42, 43]. Clearly, the present heat-capacity simulation needs only two bands to perfectly describe $C_e(T)$ for $T < T_c$, which is consistent with the BQPI/STM [8] and one of the ARPES measurements [13]. Further, the absence of a sizable $\gamma_0$ excludes the possibility of an additional electron band, either gapless or with a very small gap, with any sizeable weight [44]. Our analysis of the quantum-oscillation data [45], which yields $\gamma_n$ of 5.7 and 9.4 mJ mol$^{-1}$ K$^{-2}$ in two- and three-band models, respectively (see Supplementary Note 5), also appears to favor a two-band scenario, since the inferred $\gamma_n$ value is closer to the measured one. The two-band value is just 17 % smaller than our direct measurement. For comparison, similar calculations in BaFe$_2$As$_2$ and KFe$_2$As$_2$ [31, 46], for which the Fermi surface is fully determined, lead to comparable deviations of about 11 and 13 %, respectively.

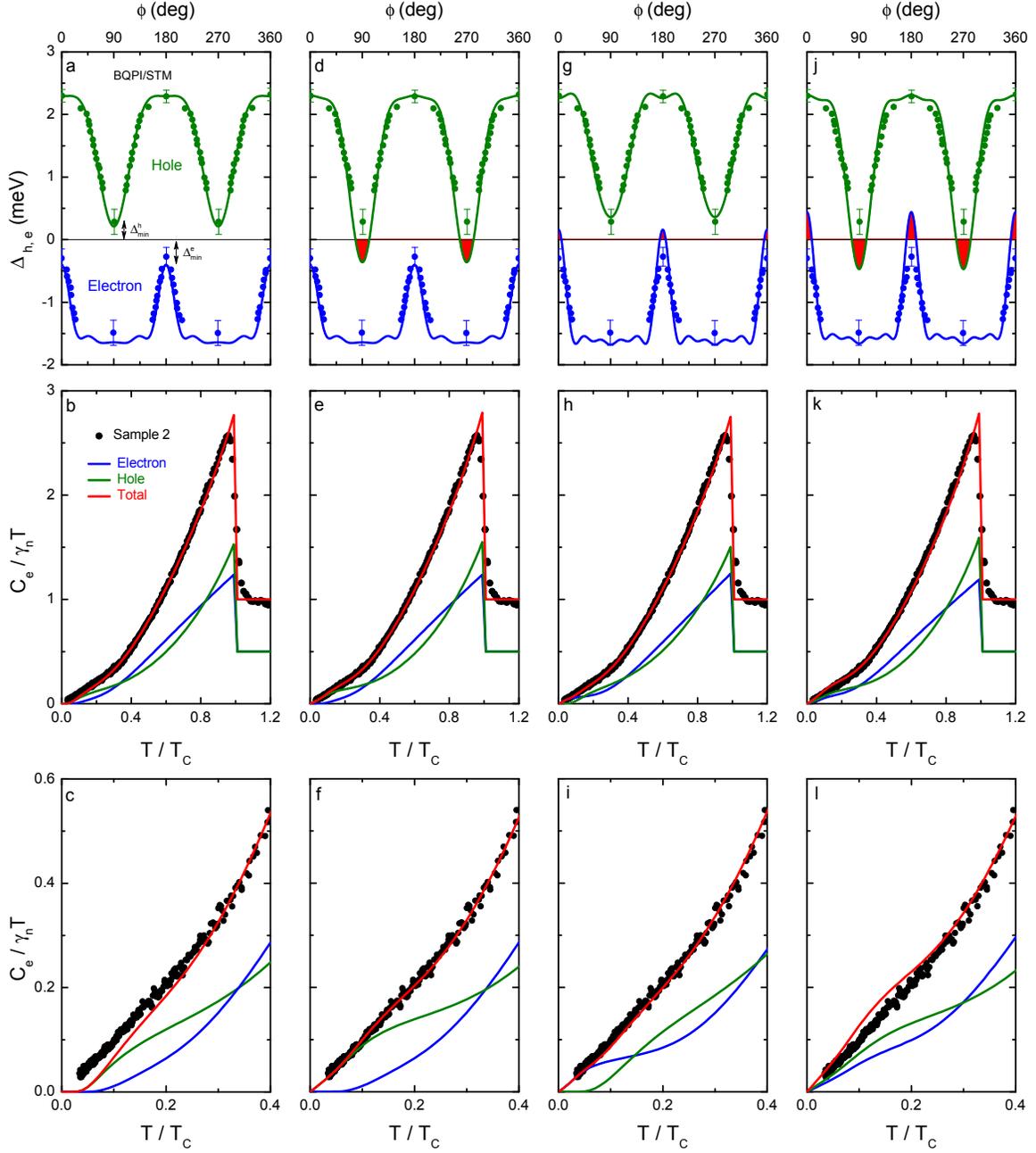

FIG. 3. (a) Nodeless angle-resolved gap structure reported by BQPI/STM [8] (green and blue symbols). (d-g-j) Nodal gap structures used for heat-capacity calculations. Solid lines are fits within the leading angular harmonic approximation. (b-e-h-k) Comparison of the calculated heat capacity (red line) to $C_e(T)$ of Sample 2. (c-f-i-l) show the low-temperature part on an enlarged scale. The green and blue lines are the individual hole and electron contributions, respectively.

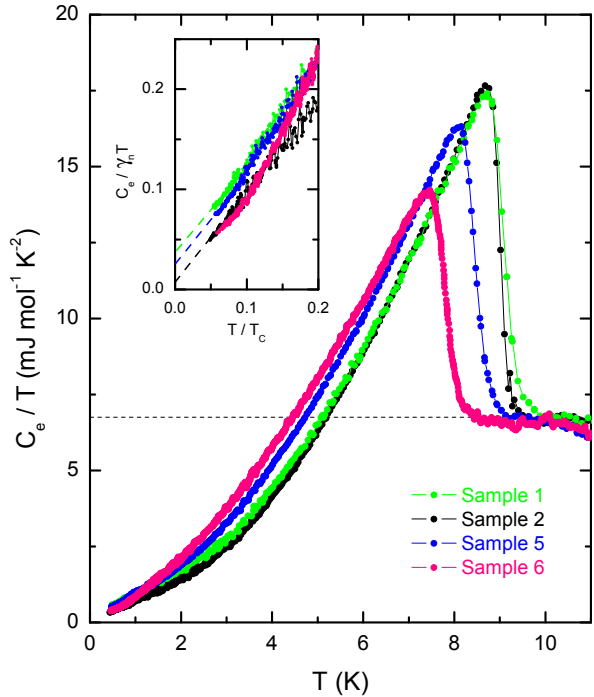

FIG. 4. (a) $C_e(T)$ of Samples 1, 2, 5 and 6 with different amounts of disorder. The inset shows the low-temperature part in normalized units on an enlarged scale.

In Refs [8, 14], it is argued that the second electron band is not observed because it may be quite incoherent as a result of strong local correlations. However, in $AFe_2As_2$ (A = K, Rb, Cs) in which the $d$ bands are closer to half-filling with strongly enhanced correlations [47, 48], one finds i) quite heavy but still coherent bands with $m^* > 20m_e$ ($m_e$ is the bare electron mass) [49] ii) a Sommerfeld coefficient exceeding 100 mJ mol$^{-1}$ K$^{-2}$ and iii) a remarkable coherence-incoherence crossover [50, 51] indicative of strong orbital-selective correlations [47, 50–52]. In FeSe, we find no evidence for such a crossover in the magnetic susceptibility (see Fig.1c) although the resistivity exhibits a broad maximum around 400 K. [53] Instead, the susceptibility of FeSe increases linearly from $T > T_s$ to at least 700 K with the same slope as $BaFe_2As_2$ [25]. Thus, the scenario of a completely incoherent second electron band appears rather unlikely and a better understanding of the nematic state is required to explain the missing electron pocket.

In conclusion, we have reported a detailed study of the thermodynamic properties of FeSe single crystals. We consistently find a linear electronic specific heat $C_e/T$ for $T << T_c$ indicative of the bulk existence of line nodes. This nodal behavior is found to be quite robust and persists in samples with significantly depressed $T_c$. Our results are consistent with a two-band scenario $i.e.$ a

Fermi surface that consists of one hole and one electron pockets only. The fate of the theoretically predicted second electron pocket remains unclear and must be related to the elusive nature of the nematic state.

*Acknowledgments.* We thank Laura Fanfarillo, Rafael Fernandes, Andrey Chubukov, Christophe Marcenat, Thierry Klein, Peter Hirschfeld, Andreas Kreisel, Brian Andersen, Matthew Watson, Paul Wiecki and Antony Carrington for fruitful discussions. A.-A.H acknowledges the financial support of the Oxford Quantum Materials Platform Grant (EP/M020517/1). The contribution from M.M. was supported by the Karlsruhe Nano Micro Facility (KNMF).